\documentclass [12pt]{iopart}
\usepackage {iopams}
\usepackage{setstack}
\usepackage{amstext}
\usepackage{amsfonts}
\usepackage{graphicx}
\usepackage{subfigure}
\usepackage{array}
\usepackage{appendix}
\usepackage{mathrsfs}
\usepackage{color}
\usepackage{makecell}
\usepackage{adjustbox,lipsum}
\usepackage{upgreek}
\usepackage{algorithm}
\usepackage{algpseudocode}
\usepackage{makecell}
\usepackage{adjustbox,lipsum}
\usepackage{dsfont}
\usepackage{amsthm}

\begin{document}
\title[]{Fundamental limits to contrast reversal of survival probability correlations}

\author{Kyoungho~Cho$^{1,2}$, Jeongho~Bang$^{2}$}

\address{$^1$ Department of Statistics and Data Science, Yonsei University, Seoul 03722, Republic of Korea}
\address{$^2$ Institute for Convergence Research and Education in Advanced Technology, Yonsei University, Seoul 03722, Republic of Korea}

\vspace{10pt}

\begin{indented}
\item Correspondence and requests for materials should be addressed to J.B.
\end{indented}

\ead{\mailto{jbang@yonsei.ac.kr}}

\begin{abstract}
In measurement design, it is common to engineer \emph{anti-contrast} readouts---two measurements that respond as differently as possible to the same inputs so that contributions that affect both readouts in the same way are suppressed. To assess the fundamental scope of this strategy in unitary dynamics, we ask whether two evolutions can be made uniformly opposite over a broad input ensemble, or whether quantum mechanics imposes a structural limit on such opposition. We address this by treating survival probability as a random variable on projective state space and adopting the Pearson correlation coefficient as a device-agnostic measure of global opposition between two evolutions. Within this framework we establish the following theorem: For any nontrivial pair of unitaries, survival probability maps cannot be point-wise complementary correlation on the entire state space. Consequently, the mathematical lower edge of the correlation bound is not physically attainable, which we interpret as a unitary-geometric floor on anti-contrast, independent of hardware specifics and noise models. We make this floor explicit in realizable settings. In a single-qubit Bloch-sphere Ramsey model, a closed-form relation shows that a residual shared (channel-symmetric) component persists even under nominally optimal tuning. In higher dimensions, Haar/design moment identities reduce ensemble means and covariances of survival probability to a small set of unitary invariants, yielding the same conclusion irrespective of implementation details. Taken together, these results provide a model-independent criterion for what anti-contrast can and cannot achieve in unitary sensing protocols.
\end{abstract}

\maketitle

\newcommand{\bra}[1]{\left<#1\right|}
\newcommand{\ket}[1]{\left|#1\right>}
\newcommand{\abs}[1]{\bigl|#1\bigr|}
\newcommand{\norm}[1]{\left\lVert#1\right\rVert}
\newcommand{\expt}[1]{\langle #1 \rangle}
\newcommand{\braket}[2]{\left<{#1}|{#2}\right>}
\newcommand{\ketbra}[2]{\ket{#1}\!\!\bra{#2}}

\newcommand{\identity}{1\!\!1}

\newtheorem{theorem}{Theorem}
\newtheorem{proposition}{Proposition}
\newtheorem{lemma}{Lemma}
\newtheorem{corollary}{Corollary}
\newtheorem{definition}{Definition}
\newtheorem{remark}{Remark}

\newcommand{\E}{\mathbb{E}}
\newcommand{\Var}{\mathrm{Var}}
\newcommand{\Cov}{\mathrm{Cov}}
\newcommand{\SU}{\mathrm{SU}}
\newcommand{\U}{\mathrm{U}}
\newcommand{\FS}{\mathrm{FS}}
\newcommand{\Id}{\mathbb{I}}
\newcommand{\PCC}{\rho} 

\section{Introduction}\label{Sec:1}

Measurement inherently disturbs the system and, as a cost, the disturbance‑induced systematic errors follow. In this context, a widely adopted strategy is a two-channel difference readout (often termed an anti-contrast, or contrast-reversal, readout)\footnote{Here and throughout, we use “anti-contrast” (contrast reversal) as an operational shorthand for a two-channel difference readout: two detector outputs are engineered to change in opposite directions under the same change in the underlying input parameter, so that a (possibly weighted) subtraction suppresses contributions that appear similarly in both channels.}, in which two detector outputs are engineered to respond oppositely to the same input and are then combined (typically by subtraction) to suppress contributions that appear similarly in both outputs~\cite{Doebelin2004,Jin2015}. For example, in some scenarios of interferometry, echo protocols, and gate benchmarking, the idea of forming a difference signal (a weighted subtraction of two readouts) is powerful, and it has in fact led to performance improvements across diverse measurement design platforms~\cite{Macri2016,Zhou2023,Gould2024mqm}. However, behind this intuition there still remains a question to be considered. No matter how one changes device details, when one scans a broad ensemble of the prepared signals, can two quantum evolutions be made everywhere exactly the opposite of each other? In other words, beyond device specifics and noise models, does there exist a structural limit inherent to quantum mechanics? Answering it is important in that, at the theoretical level, it provides a common language for fair comparison and calibration that strips away input‑specific or device‑specific details. Moreover, it clarifies the fundamental extent to which a two-channel difference readout can suppress contributions that appear similarly in both channels, and whether the experimental aspiration of ``perfect cancellation'' should be viewed as an achievable ideal or, instead, as an optimization problem with an irreducible floor.

From a quantum-information or quantum-metrological perspective, introducing two readout channels is not a fundamental requirement, but a practical strategy for constructing the estimators which are robust to some noisy variations~\cite{Suzuki2020}. Importantly, the ``two channels'' need not be two distinct physical detectors: they can correspond to two output ports, two measurement settings, or two control sequences applied to the same apparatus. The operational motivation is that many experimental imperfections enter both readouts in a similar way (e.g., offsets, slow drifts, or calibration-dependent gains), whereas the parameter-dependent response can be engineered to change with opposite sign in the two channels~\cite{Robinson2012,Jin2015}. Thus, by forming a (possibly weighted) difference of the two readouts, we can suppress the shared component while retaining---often amplifying---the parameter-dependent contrast. In this sense, the ``maximally anti-correlated'' readouts can represent an idealized target: if two readout maps are point-wise complementary over the relevant input ensemble, a linear recombination can, in principle, isolate the contrast while canceling the shared component. The central question we address is how far such global opposition can be pushed when the readouts are the probabilities generated by {\em quantum unitary} dynamics, independent of any specific hardware implementation or noise model.

Motivated by this two-channel difference-readout viewpoint, our study takes the survival probability---the probability of returning to the input state after a unitary evolution~\cite{Fonda1978,Peres1984}---as a basic random variable on state space, and formalizes the global opposition of two evolutions by means of the Pearson correlation coefficient (PCC)~\cite{Pearson1895,CasellaBerger2002}. This indicator, PCC, is insensitive to mere rescaling of contrast and does not depend on detailed assumptions about devices or models. Within this framework we examine whether the long‑pursued experimental ideal---``wherever one is bright, the other is dark''---can be physically realized, namely whether the survival probability maps can form a point-wise complementary relation over the entire state space. The core message obtained through this study is simple and conclusive. Even under only the minimal structure of Hilbert space, there exists a universal upper bound on the global correlation of two random variables, and this bound is attainable only when one variable is an affine transform of the other~\cite{Witsenhausen1975,Anantharam2013}. However, by specializing this to pairs of the survival probabilities, we prove that the ideal point-wise complementary relation is fundamentally excluded. In other words, a correlation boundary permitted by quantum theory becomes a point that is not attainable. To the best of our knowledge, this is the first result that generally, i.e., without protocol‑specific assumptions, excludes a perfectly opposite correlation based on the survival probability in finite‑dimensional unitary dynamics.

This structure appears vividly in concrete platforms. In a single‑qubit Bloch‑sphere Ramsey model~\cite{Clemmen2016,Roszak2020}, the geometry between the two unitary control axes effectively determines the magnitude of the global correlation, and the pulse strength changes only the contrast without changing the structure of the interference pattern. Consequently, even when the parameters that implement the output pattern by quantum evolution are arranged optimally, a residual shared component of the output remains and cannot be fully canceled. As one goes to higher dimensions, through Haar moment identities showing that the ensemble mean and covariance of the survival probability are summarized by a small number of unitary invariants~\cite{Collins2006,Matsumoto2023}, the same conclusion is drawn independent of device details. In the short‑time regime, the leading fluctuation of the survival probability is linked to state‑dependent energy variances; to reach an ideal opposite pattern would require very strong constraints among those variances, but these cannot be satisfied for nontrivial Hamiltonians. This limitation does not come from a particular parameter combination or circuit compilation, but from the unitary‑geometric fact of how a unitary map places bright and dark over state space.

From this, a practical intuition becomes clear. Within our framework, what the experiments can actually adjust to reduce the global correlation lies within the unitary geometry of the quantum system. By separating control axes, choosing interrogation times appropriately, and assembling input ensembles close to projective designs, one can meaningfully reduce the overlap of the survival probability patterns~\cite{Nakata2021,Oviedo2024,Iosue2024}. On the other hand, in the quantum model considered here, making a perfect bright‑versus‑dark tiling of the entire state space is fundamentally impossible. In addition, even if one forms an optimally weighted contrast by taking two survival probabilities as features, a floor of variance remains that does not disappear completely. Our study framework can serve as a useful metric for fair cross‑platform comparison and for pushing experiments as close as possible to the boundary allowed by unitary geometry. 

\section{Pearson Correlation Coefficient and Cauchy-Schwarz bound}\label{sec:2}

In this section we recall the Pearson correlation coefficient (PCC), emphasize its Cauchy-Schwarz bound as a fundamental structural fact. Let $X$ and $Y$ be real random variables on a probability space. The expectation and variance are
\begin{eqnarray}
\E[X] = \int_\Omega X\,d\mu, \quad \Var(X) &=& \E \left[ (X-\E[X])^2 \right],
\end{eqnarray}
where $\mu$ is a probability measure. Then, let us recall the notion of covariance, which measures the joint variability of two random variables~\cite{CasellaBerger2002},
\begin{eqnarray}
\Cov(X,Y) &=& \E \left[ (X-\E[X])(Y-\E[Y]) \right].
\end{eqnarray}
This covariance tells us whether the two variables increase together (positive covariance), vary oppositely (negative covariance), or are unrelated in their fluctuations (zero covariance). Since the covariance depends on the units and scaling of $X$ and $Y$, it is often more convenient to normalize; it leads to the definition of the PCC as~\cite{Pearson1895,CasellaBerger2002}
\begin{eqnarray}
P(X,Y) := \frac{\Cov(X,Y)}{\sqrt{\Var(X)} \sqrt{\Var(Y)}} = \frac{\Cov(X,Y)}{\Delta_X \Delta_Y},
\label{eq:PCC}
\end{eqnarray}
where we use the following notation: $\Delta_A^2:=\Var(A)$ and $\Delta_A:=\sqrt{\Var(A)}$. The PCC satisfies the following properties: (i) Symmetry: $P(X,Y)=P(Y,X)$. (ii) Affine behavior: For $X'=aX+b$, $Y'=cY+d$ with $a,c \neq 0$, we have $P(X',Y') = \mathrm{sgn}(ac) P(X,Y)$; the translations and positive rescalings leave $P$ invariant, while negating exactly one variable flips its sign. (iii) Independence: If $X$ and $Y$ are independent (with finite second moments), then $P(X,Y)=0$; the converse need not hold. Here, note also that $P(X,Y)$ is undefined if either variance vanishes (degenerate readout) and $P=0$ does not imply independence, i.e., nonlinear dependencies can persist even when linear correlation vanishes. 

Now, we provide a more special property of PCC as a theorem:
\begin{theorem}[Cauchy-Schwarz bound for PCC]
\label{thm:cs-pcc}
Let $X, Y \in L^2(\mu)$ with $\Delta_X, \Delta_Y>0$. Then,
\begin{eqnarray}
-1 \le P(X,Y) \le 1,
\label{eq:CS-PCC}
\end{eqnarray}
with equality if and only if $Y=aX+b$ almost surely for some $a \neq 0$ and $b \in \mathbb{R}$. Here, $L^2(\mu)$ denotes the real Hilbert space of (equivalence classes of) square‑integrable random variables, i.e., $L^2(\mu):=\{Z:\Omega\to\mathbb{R}\ \mid\ \E[Z^2]<\infty\}$, equipped with the inner product $\langle A,B \rangle := \E[AB]$ and norm $\|A\| := \sqrt{\E[A^2]}$. The term ``almost surely''  means ``except on a $\mu$‑null set.''
\end{theorem}

\begin{proof}
The proof is simple and concise. Firstly, let $X_c := X - \E[X]$ and $Y_c := Y - \E[Y]$. With the $L^2(\mu)$ inner product $\langle A,B \rangle := \E[AB]$, we can express 
\begin{eqnarray}
P(X,Y)=\frac{\langle X_c,Y_c\rangle}{\|X_c\|\,\|Y_c\|}.
\end{eqnarray}
Cauchy-Schwarz inequality gives $\abs{\langle X_c,Y_c\rangle} \le \|X_c\| \|Y_c\|$, hence $\abs{P} \le 1$ holds~\cite{Wackerly08}. Equality in the Cauchy-Schwarz inequality holds if and only if $Y_c=\lambda X_c$ almost surely for some $\lambda \in \mathbb{R}$, i.e., $Y=\lambda X + b$ with $b=\E[Y] - \lambda \E[X]$ and $\lambda \neq 0$ because $\Delta_Y > 0$.
\end{proof}

In the $L^2$ geometry induced by $\langle\cdot,\cdot\rangle$, the normalized fluctuations $X_c/\Delta_X$ and $Y_c/\Delta_Y$ are unit vectors, and $P(X,Y)$ is their cosine. The universal cap $\abs{P} \le 1$ therefore states that the directional overlap of the two fluctuation patterns is bounded by unity.  As $\abs{P} \to 1$, the joint scatter of $(X,Y)$ collapses onto a line: one readout carries no independent linear information beyond the other (complete redundancy). As $\abs{P} \to 0$, the fluctuation directions become nearly orthogonal: the readouts encode complementary linear information~\cite{Abdi2010}.

\section{Survival Probability Correlation Model and Impossibility of Perfect Anti-Correlation}\label{sec:3}

We instantiate the PCC framework of Sec.~\ref{sec:2} in a quantum setting. Let $\mathcal{H} \cong \mathbb{C}^d$ be a $d$-dimensional Hilbert space, and let $\hat{U}_1, \hat{U}_2 \in \mathrm{U}(d)$ be two (fixed) unitary operations~\footnote{Global phases are immaterial for all quantities below; one may take $\hat{U}_\alpha \in \mathrm{SU}(d)$ without loss of generality.}. For a pure state $\ket{\psi} \in \mathcal{H}$, we define the survival probability random variables~\cite{Peres1984}:
\begin{eqnarray}
X_j(\ket{\psi}) := \abs{\bra{\psi}\hat{U}_j\ket{\psi}}^2 \quad (j=1,2).
\label{eq:rand_Xj}
\end{eqnarray}
We average over the pure states with respect to the measure $d\psi$ on projective space (the unique unitarily invariant probability measure); we here abbreviate $\E_\psi[\cdot] := \int (\cdot) d\psi$.
Thus, the variance is
\begin{eqnarray}
\Delta_j^2 := \E_\psi[X_j(\ket{\psi})^2] - \E_\psi[X_j(\ket{\psi})]^2.
\end{eqnarray}
The PCC can be defined as
\begin{eqnarray}
P(\hat{U}_1, \hat{U}_2) := \frac{\Cov(X_1, X_2)}{\Delta_1\Delta_2},
\end{eqnarray}
where 
\begin{eqnarray}
\Cov(X_1, X_2) = \E_\psi[X_1 X_2] - \E_\psi[X_1] \E_\psi[X_2].
\end{eqnarray}
Here, two structural remarks are used repeatedly: (i) $X_{\hat{V} \hat{U}_j \hat{V}^\dagger}(\hat{V}\ket{\psi})=X_{\hat{U}_j}(\ket{\psi})$ for any unitary $\hat{V}$ (unitary-conjugation covariance), and (ii) $X_{e^{i\phi} \hat{U}_j}\equiv X_{\hat{U}_j}$ (global-phase invariance). Thus, $P(\hat{U}_1, \hat{U}_2)$ depends only on the relative unitary data of the pair $(\hat{U}_1, \hat{U}_2)$. Note further that for any unitary $\hat{U}$, $X_{\hat{U}}(\ket{\psi}) \in [0,1]$ and $X_{\hat{U}}(\ket{\psi})=1$ whenever $\ket{\psi}$ is an eigenstate of $\hat{U}$.
If $\hat{U}$ is trivial (a global phase), then $X_{\hat{U}} = 1$ and $\Delta_{\hat{U}}=0$, in which case $P$ is undefined. Thus, we exclude such degeneracies by assuming $\Delta_1,\Delta_2>0$.

Then, we formalize in our setting the well-known statement that there are no two nontrivial unitaries that map every input state to mutually orthogonal outputs.

\begin{theorem}[No universal quantum inversion~\cite{Buzek99,Rungta2001,Bang2012}]
\label{thm:no-universal-inversion}
There are no nontrivial $\hat{U}_1, \hat{U}_2 \in \mathrm{U}(d)$ such that
\begin{eqnarray}
\bra{\psi} \hat{U}_2^\dagger \hat{U}_1 \ket{\psi} = 0~\text{for all states}~\ket{\psi} \in \mathcal{H}.
\label{eq:univ-orth}
\end{eqnarray}
Equivalently, there are no nontrivial $\hat{U}_1, \hat{U}_2$ for which $\hat{U}_1\ket{\psi} \perp \hat{U}_2\ket{\psi}$ holds for every $\ket{\psi}$.
\end{theorem}

\begin{proof}
Let $\hat{M} := \hat{U}_2^\dagger \hat{U}_1$. Assumption Eq.~(\ref{eq:univ-orth}) says $\bra{\psi}\hat{M}\ket{\psi}=0$ for all $\ket{\psi}$. Given $\ket{\psi}, \ket{\phi} \in \mathcal{H}$, let us define the auxiliary (unnormalized) kets as $\ket{\chi_k} := \ket{\psi} + i^k \ket{\phi}$. Then, by the complex polarization identity in Hilbert spaces,
\begin{eqnarray}
\bra{\phi}\hat{M}\ket{\psi} = \frac{1}{4} \sum_{k=0}^3 i^k \bra{\chi_k} \hat{M} \ket{\chi_k} = 0,
\end{eqnarray}
so $\bra{\phi}\hat{M}\ket{\psi}=0$ for all $\ket{\phi}$ and $\ket{\psi}$, whence $\hat{M}=\hat{U}_2^\dagger \hat{U}_1$ is null---which is contradiction.
\end{proof}

We now establish the link between the anti-correlation of the random variables $X_j(\ket{\psi})$ and the no universal inversion theorem~\cite{Van2005}. Assume, for contradiction, that the following point-wise complementarity ansatz holds:
\begin{eqnarray}
X_1(\ket{\psi}) + X_2(\ket{\psi}) = 1~\text{for all}~\ket{\psi}.
\label{eq:sum-to-one}
\end{eqnarray}
Let $\{\ket{e_k}\}_{k=1}^d$ be an eigenbasis of $\hat{U}_1$, i.e., $\hat{U}_1\ket{e_k}=e^{i\theta_k}\ket{e_k}$. Plugging $\ket{\psi}=\ket{e_k}$ into Eq.~(\ref{eq:sum-to-one}), we obtain $X_2(\ket{e_k})=0$, hence $\bra{e_k}\hat{U}_2\ket{e_k}=0$ for all $k$, and therefore $\tr(\hat{U}_2)=0$. By the same reasoning with $\hat{U}_1$ and $\hat{U}_2$ interchanged, we also obtain $\tr(\hat{U}_1)=0$.

Now average Eq.~(\ref{eq:sum-to-one}) over the unitarily invariant measure $d\psi$ on the projective space. Using the standard projective $2$-design identity (see~{\bf \ref{appendix:A}}), $\mathbb{E}_\psi[X_{\hat{U}}] = \frac{d+\abs{\tr(\hat{U})}^2}{d(d+1)}$, we get $\mathbb{E}_\psi[X_1]=\mathbb{E}_\psi[X_2]=\frac{1}{d+1}$ and thus $\mathbb{E}_\psi[X_1+X_2]=\frac{2}{d+1} \neq 1$ for $d \ge 2$. This contradicts Eq.~(\ref{eq:sum-to-one}). Consequently, the point-wise complementarity ansatz is impossible for any nontrivial unitary pair.

\begin{corollary}[No perfect complement]
\label{cor:no-perfect-complement}
There are no nontrivial unitary pair $(\hat{U}_1, \hat{U}_2)$, satisfying 
\begin{eqnarray}
X_1(\ket{\psi})+X_2(\ket{\psi})=1 \quad \text{for all $\ket{\psi}$}.
\end{eqnarray}
\end{corollary}
It directly leads to the consequences for PCC. Recall from {\bf Theorem~\ref{thm:cs-pcc}} that $P = -1$ holds iff the two random variables are affinely related almost surely, i.e.,
\begin{eqnarray}
X_2 = a X_1 + b \quad (a < 0,~b \in \mathbb{R}),
\label{eq:affine-saturation}
\end{eqnarray}
with nonzero variances. In our quantum setting, $X_j(\ket{\psi}) \in [0,1]$ and $X_j(\ket{\psi})=1$ occurs on eigenstates of $\hat{U}_j$ ($j=1,2$). A direct re-centering and re-scaling of Eq.~(\ref{eq:affine-saturation}) produces a monotone affine encoding
\begin{eqnarray}
\widetilde{X}_2(\ket{\psi}) := \frac{X_2(\ket{\psi}) - b}{-a} + 1.
\end{eqnarray}
Here $\tilde{X}_2$ is introduced only as an affine reparameterization (i.e., a classical calibration) of the second readout implied by $P=-1$; we do not interpret $\tilde{X}_2$ itself as a survival probability of any unitary. Then, we have
\begin{eqnarray}
X_1(\ket{\psi}) + \widetilde{X}_2(\ket{\psi}) = 1.
\label{eq:affine-to-complement}
\end{eqnarray}
Therefore, if $P(\hat{U}_1, \hat{U}_2)=-1$ is achievable with the survival probabilities, then a mere affine re-labeling of the second readout (which does not alter the $L^2$-geometry behind {\bf Theorem~\ref{thm:cs-pcc}}) would enforce the complementary identity in Eq.~(\ref{eq:affine-to-complement}). However, {\bf Corollary~\ref{cor:no-perfect-complement}} shows that no pair of unitaries can realize this point-wise complementarity; hence, we establish a theorem as follow:
\begin{theorem}[Impossibility of $P=-1$ for survival probabilities]
\label{thm:no-minus-one}
For two nontrivial unitary operations $\hat{U}_1, \hat{U}_2 \in \mathrm{U}(d)$, the perfect anti-correlation of the random variables $X_j(\ket{\psi}) = \bigl| \bra{\psi}\hat{U}_j\ket{\psi} \bigr|^2$ ($j=1,2$) cannot be attained; equivalently, $P(\hat{U}_1, \hat{U}_2) = -1$ is not reachable.
\end{theorem}

The unattainability of $P=-1$ for survival probabilities means that, within the present (a-sort-of) gate-comparison model, one cannot realize a pair of readouts whose linear fluctuations are \emph{perfectly} anti-aligned across the entire input-state ensemble. Metrologically, the ``contrast-only'' limit has no physical realization here: some residual shared sensitivity across the two readouts remains inescapable.

\section{Survival Probability in Practice: Bloch-sphere Ramsey Interferometry and Loschmidt Echo}\label{sec:4}

\subsection{Single-qubit Bloch-sphere Ramsey interferometry} 

\begin{figure}[t]
\centering
\includegraphics[width=0.65\textwidth]{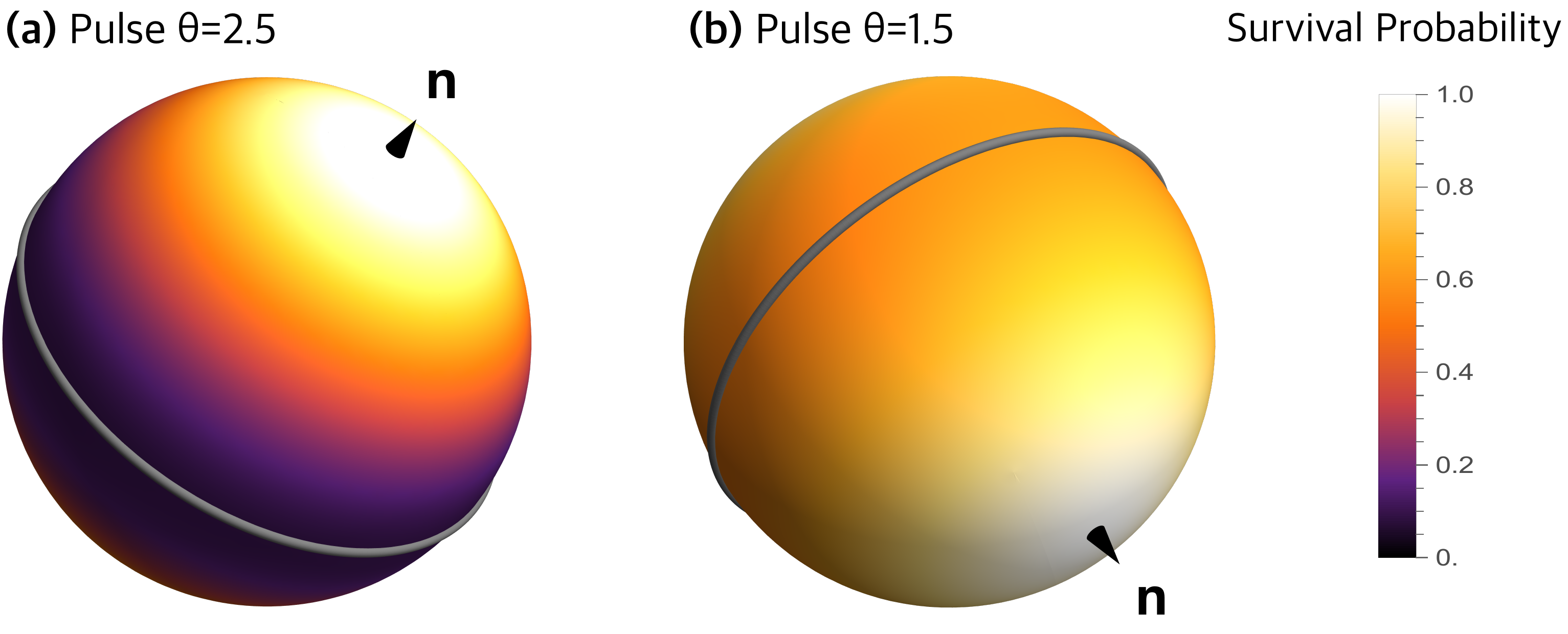}
\caption{Survival probability fringe patterns on the single-qubit Bloch-sphere Ramsey interferometry model. For a single-qubit unitary $\hat{U}(\theta, \mathbf{n})=\exp\left(-\frac{i}{2} \theta \mathbf{n} \cdot \boldsymbol{\sigma} \right)$, the color encodes $X_{\hat{U}}$ in Eq.~(\ref{eq:qubit-fidelity-formula}). Here, it is observed that the two bright caps are centered at $\pm\mathbf n$ and the dark belt follows the great circle $\mathbf r\!\cdot\!\mathbf n=0$. (a) $\theta=2.5$ (high contrast): the bright/dark separation is strong because the amplitude $A=\sin^2{(\frac{\theta}{2})}$ is large. (b) $\theta=1.5$ (lower contrast): the cap/belt fringe geometry is unchanged---only the contrast amplitude is reduced. Thus, $\mathbf{n}$ fixes the location of the bright/dark regions, while $\theta$ only rescales their contrast.}
\label{fig:BS_Ramsey} 
\end{figure}

Consider a qubit subject to two coherent controls $\hat{U}_j=\exp\left(-\frac{i}{2} \theta_j \mathbf{n}_j \cdot \boldsymbol{\sigma} \right)$ with rotation angle $\theta_j \in (0,2\pi)$ and unit Bloch axes $\mathbf{n}_j \in \mathbb{R}^3$ ($j=1,2$). For a pure state with Bloch vector $\mathbf{r} \in \mathbb{S}^2$ (i.e., $\hat{\rho}=\frac{1}{2}(\hat{\mathds{1}} + \mathbf{r} \cdot \boldsymbol{\sigma})$),
\begin{eqnarray}
 X_{\hat{U}_j} = \abs{\tr(\hat{\rho}\hat{U}_j)}^2  = 1 - A_j \left( 1 - (\mathbf{n}_j \cdot \mathbf{r})^2 \right),
\label{eq:qubit-fidelity-formula}
\end{eqnarray}
where $\boldsymbol{\sigma}=(\hat{\sigma}_x, \hat{\sigma}_y, \hat{\sigma}_z)^T$ and $A_j$ is the amplitude, given by $A_j = \sin^2{(\frac{\theta_j}{2})}$. On the Bloch sphere this looks as a fringe pattern drawn by the axis $\mathbf{n}_j$ and modulated by the angle $\theta_j$, as in Fig.~\ref{fig:BS_Ramsey}: Here we can observe the two bright caps centered at $\pm\mathbf{n}_j$ (high survival probability) and a dark band around the great circle orthogonal to $\mathbf{n}_j$ (low survival probability). The angle $\theta_j$ changes the contrast (i.e., how strongly bright and dark separate), but it does not move the pattern on the sphere; only the axis $\mathbf{n}_j$ sets where the bright and dark directions live.

Averaging over all input states with the Fubini-Study measure corresponds to the uniform measure on $\mathbb{S}^2$, for which (see~{\bf \ref{appendix:A}} or Ref.~\cite{Collins2006,Dankert2009})
\begin{eqnarray}
\E_{\mathbf{r}}[(\mathbf{n} \cdot \mathbf{r})^2]=\frac{1}{3}, \qquad \E_{\mathbf{r}}[(\mathbf{n} \cdot \mathbf{r})^4]=\frac{1}{5}.
\end{eqnarray}
A simple calculation from Eq.~(\ref{eq:qubit-fidelity-formula}) yields
\begin{eqnarray}
\E_{\mathbf{r}}[X_{\hat{U}_j}] = \frac{1}{3} + \frac{2}{3}\cos^2{\frac{\theta_j}{2}}, \qquad \Delta_j^2 = \frac{4}{45}\sin^4{\frac{\theta_j}{2}},
\label{eq:qubit-var}
\end{eqnarray}
where $\theta_j \not\equiv 0 \ \mathrm{mod} \ 2\pi$. Here, let $\gamma := \mathbf{n}_1 \cdot \mathbf{n}_2=\cos\delta$ be the cosine of the angle $\delta \in [0,\pi]$ between control axes. Then, 
\begin{eqnarray}
\E_{\mathbf{r}} \bigl[(\mathbf{n}_1 \cdot \mathbf{r})^2 (\mathbf{n}_2 \cdot \mathbf{r})^2 \bigr] = \frac{1 + 2\gamma^2}{15},
\end{eqnarray}
which directly gives
\begin{eqnarray}
\Cov(X_{\hat{U}_1}, X_{\hat{U}_2}) = \frac{2}{45} \sin^2{\frac{\theta_1}{2}} \sin^2{\frac{\theta_2}{2}} \bigl( 3\gamma^2 -1 \bigr).
  \label{eq:qubit-cov}
\end{eqnarray}
Combining Eq.~(\ref{eq:qubit-var}) and  Eq.~(\ref{eq:qubit-cov}), we obtain a simple, angle-only, formula:
\begin{proposition}[Qubit geometric correlation]
\label{thm:qubit-P-closed}
For any nontrivial single-qubit rotations $\hat{U}_1,\hat{U}_2$, the PCC of survival probabilities is
\begin{eqnarray}
P(\hat{U}_1, \hat{U}_2) = \frac{\Cov(X_{\hat{U}_1}, X_{\hat{U}_2})}{\Delta_1, \Delta_2} = \frac{3\cos^2\delta-1}{2},
\label{eq:qubit-P-closed}
\end{eqnarray}
where $\delta=\angle(\mathbf{n}_1,\mathbf{n}_2)$. Note here that $P$ is independent of the rotations $\theta_1, \theta_2$, and it obeys the sharp bounds
\begin{eqnarray}
-\frac{1}{2} \le P(\hat{U}_1, \hat{U}_2) \le 1,
\end{eqnarray}
with
\begin{eqnarray}
  \begin{array}{ll}
    P=1 & \Leftrightarrow\ \delta=0\ \text{or}\ \pi\ \ (\mathbf{n}_1\parallel\mathbf{n}_2),\\
    P=-\frac{1}{2} & \Leftrightarrow\ \delta=\frac{\pi}{2}\ \ (\mathbf{n}_1\perp\mathbf{n}_2).
  \end{array}
\label{eq:qubit-P-bounds}
\end{eqnarray}
\end{proposition}

\begin{figure}[t]  
\centering
\includegraphics[width=0.70\textwidth]{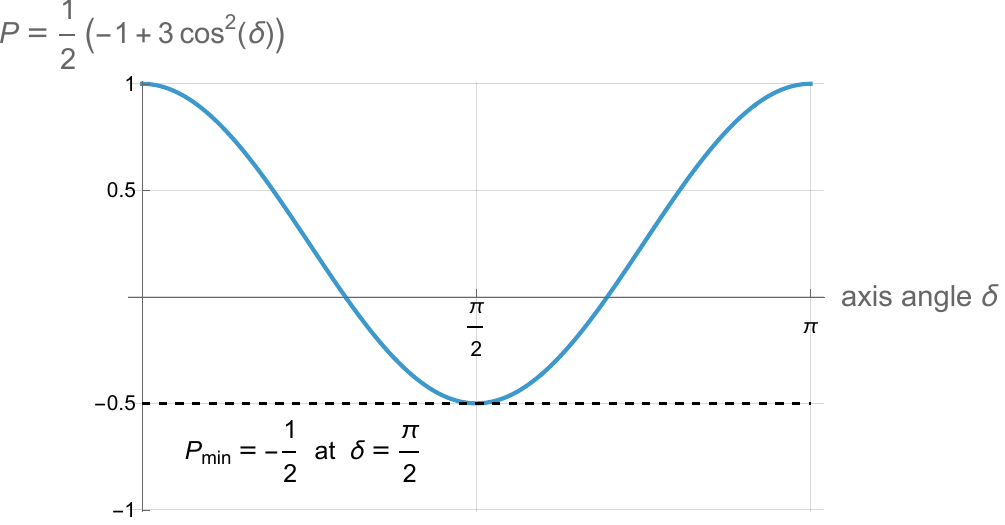}
\caption{The qubit survival probability correlation $P$ vs $\delta$ between Ramsey axes. Perfect anti-correlation $P=-1$ is forbidden; the best anti-correlation is $P=-\frac{1}{2}$.}
\label{fig:P_delta}
\end{figure}

\begin{proof}
Eqs.~(\ref{eq:qubit-var})--(\ref{eq:qubit-cov}) give 
\begin{eqnarray}
\Delta_j = \frac{2}{\sqrt{45}}\sin^2{\frac{\theta_j}{2}}, \quad \Cov(X_{\hat{U}_1}, X_{\hat{U}_2}) = \frac{2}{45}\sin^2{\frac{\theta_1}{2}}\sin^2{\frac{\theta_2}{2}} \left( 3\gamma^2-1 \right).
\end{eqnarray}
Divide to obtain Eq.~(\ref{eq:qubit-P-closed}); then, Eq.~(\ref{eq:qubit-P-bounds}) follows from $\gamma^2 = \cos^2\delta \in [0,1]$.
\end{proof}

The physical interpretation then follows immediately from Eq.~(\ref{eq:qubit-P-closed}). Averaging over all input states, $P$ between two Bloch-sphere Ramsey survival maps quantifies how similarly the two patterns brighten and dim the same states. Note that $P$ in Eq.~(\ref{eq:qubit-P-closed}) depends only on the axis angle $\delta$, \emph{not} on the pulses $\theta_1, \theta_2$. Changing $\theta_j$ merely rescales each map’s contrast; Pearson centering and normalization remove that scale, leaving a geometric dependence on $\delta$. The key point is that the best anti‑correlation occurs at $\delta=\pi/2$ with $P_{\min}=-\frac{1}{2}$ (not $-1$) [see Fig.~\ref{fig:P_delta}]: orthogonal axes maximize the chance that a state bright for one map sits near the other’s dark belt, however the qubit geometry forbids perfect interlocking. A large ``middle'' set of states---those roughly halfway between $\mathbf{n_1}$ and $\mathbf{n_2}$---excites both sequences moderately, enforcing an irreducible shared (channel-symmetric) component. Operationally, the message is simple and robust: ``perfect anti-contrast'' in the point-wise sense does not exist for nontrivial Ramsey sequences. Even with orthogonal axes and carefully chosen phases, some residual shared sensitivity is unavoidable. The right design goal, therefore, is not to eliminate this shared component but to \emph{minimize} it (e.g., by choosing axes as close to orthogonal as constraints allow), fully aware of the geometric floor certified by Eq.~(\ref{eq:qubit-P-closed}).

In a statistical readout design, one can form the linear contrast $C_\kappa = X_1 - \kappa X_2$ from two Ramsey readouts~\cite{Len2022,Hainzer2024}. The optimal weight $\kappa^\star = \Cov(X_1,X_2)/\Delta_2^2$ minimizes the contrast variance to
\begin{eqnarray}
\min_\kappa \Var(C_\kappa)=\Delta_1^2\bigl(1-P^2\bigr).
\end{eqnarray}
Here, if $P=-1$ is achievable, this ``anti‑contrast'' channel would be noiseless; however, in our Bloch‑sphere Ramsey model does not allow $P=-1$ (its most negative value is $-\frac{1}{2}$ at $\delta=\frac{\pi}{2}$), so even at $\kappa=\kappa^\star$ a nonzero floor remains.

\subsection{Beyond qubits: Loschmidt echo} 

The survival probability $X_{\hat U}(\ket{\psi})=\abs{\bra{\psi}\hat{U}\ket{\psi}}^2$ extends to many-body Hilbert spaces, known as the \emph{Loschmidt echo}~\cite{Gorin2006}. To place the qubit model into this higher-dimensional setting, we fix a finite dimension $d$ and consider two drives of equal interrogation time $t$,
\begin{eqnarray}
\hat{U}_j=e^{-i\hat{H}_j t} \quad (j=1,2),
\end{eqnarray}
with distinct Hamiltonians $\hat H_j$. For a pure input state $\ket\psi$, the quantities $X_{\hat{U}_j}(\ket{\psi})$ measure the return under each drive, while the relative echo,
\begin{eqnarray}
X_{\mathrm{rel}}(\ket{\psi})=\abs{\bra{\psi} \hat{U}_2^\dagger\hat{U}_1 \ket{\psi}}^2
\end{eqnarray}
probes their mismatch on the same preparation. The ensemble viewpoint in Sec.~\ref{sec:3} lifts to $d>2$ by sampling the inputs $\ket{\psi}$ from a distribution absolutely continuous with respect to Haar measure on projective space (in practice, an approximate projective design suffices). In this regime the mean, variance and covariance of $X_{\hat U_j}$ admit closed forms that depend only on unitary invariants, so PCC becomes a \emph{geometric} functional of $\hat U_1$ and $\hat U_2$.

This model can be formalized with Haar-moment identities. A standard projective $2$-design calculation gives~\cite{Dankert2009}
\begin{eqnarray}
\E_\psi[X_{\hat{U}}] = \frac{\abs{\tr(\hat{U})}^2+d}{d(d+1)}.
\label{eq:bq-mean}
\end{eqnarray}
One convenient proof is to use the following:
\begin{eqnarray}
&& \int d\psi \ket{\psi}\bra{\psi}^{\otimes 2} = \frac{\hat{\mathds{1}}+\hat{S}}{d(d+1)}, \nonumber \\
&& \abs{\bra{\psi}\hat{U}\ket{\psi}}^2 = \tr \left[(\ket{\psi}\bra{\psi})^{\otimes 2}(\hat{U} \otimes \hat{U}^\dagger)\hat{S}\right],
\end{eqnarray}
where $\hat{S}$ is the swap operator: $\hat{S} = \sum_{j,k=1}^{d} \ket{jk}\bra{kj}$. The mixed fourth-order average is fixed by a projective $4$-design identity,
\begin{eqnarray}
\E_\psi[X_{\hat{U}_1} X_{\hat{U}_2}]=\sum_{l}c_l(d) \, \mathcal{I}_l(\hat{U}_1,\hat{U}_2),
\label{eq:bq-fourth}
\end{eqnarray}
where the invariants $\mathcal{I}_l$ can be chosen from $\abs{\tr(\hat{U}_1)}$, $\abs{\tr(\hat{U}_2)}$, $\abs{\tr(\hat{U}_1\hat{U}_2)}$, $\abs{\tr(\hat{U}_1\hat{U}_2^\dagger)}$ and $\tr(\hat{U}_1\hat{U}_2\hat{U}_1^\dagger\hat{U}_2^\dagger)$, with coefficients $c_l(d)$ listed in~{\bf \ref{appendix:A}}. Since $\Delta_j^2$ and $\Cov(X_{\hat{U}_1}, X_{\hat{U}_2})$ are linear combinations of such averages, $P$ depends only on these invariants and, in particular, is insensitive to the choice of an approximate design as long as the design reproduces the relevant moments. In the commuting case (i.e., $\bigl[ \hat{U}_1, \hat{U}_2 \bigr]=0$) the two unitaries are jointly diagonalizable and necessarily share the bright caps in projective space, so perfect complementarity is impossible. In the noncommuting case the commutator-type invariant $\tr(\hat{U}_1 \hat{U}_2 \hat{U}_1^\dagger \hat{U}_2^\dagger)$ upper bounds how negative the covariance can be, and the $4$-design identity provides explicit expressions for the relevant moments (and hence for $\mathrm{Var}$ and $\mathrm{Cov}$) in terms of a small set of unitary invariants (as in {\bf \ref{appendix:A}}); in particular, consistent with the general no-go result of {\bf Theorem~\ref{thm:no-minus-one}}, these expressions never attain $P=-1$ for any nontrivial unitary pair in finite dimension. Thus, the qubit obstruction to perfect anti-contrast persists in higher dimensions as a consequence of unitary geometry.

This obstruction appears at short times. A second-order expansion of $\bra{\psi} e^{-i\hat{H}_j t} \ket{\psi}$ gives
\begin{eqnarray}
\bra{\psi} e^{-i\hat{H}_j t} \ket{\psi} = 1 - i t \expt{\hat{H}_j}_\psi - \frac{t^2}{2} \expt{\hat{H}_j^2}_\psi + O(t^3),
\end{eqnarray}
hence, the survival probability is
\begin{eqnarray}
X_{\hat{U}_j}(\ket{\psi}) &=& 1 - t^2\left( \expt{\hat{H}_j^2}_\psi - \expt{\hat{H}_j}_\psi^2 \right) + O(t^4), \nonumber \\
  &=& 1 - t^2\,\Var_\psi(\hat{H}_j) + O(t^4).
\label{eq:bq-shorttime}
\end{eqnarray}
Over the input ensemble the centered fluctuations obey
\begin{eqnarray}
X_{\hat U_j}-\E[X_{\hat U_j}]\ = -t^2\Big(\Var_\psi(\hat H_j)-\E[\Var_\psi(\hat H_j)]\Big)+O(t^4),
\end{eqnarray}
so to leading nontrivial order
\begin{eqnarray}
&& \Delta_j^2 = t^4\,\Var{\bigl(\Var_\psi(\hat{H}_j)\bigr)} + O(t^6), \nonumber \\
&& \Cov(X_{\hat{U}_1}, X_{\hat{U}_2}) = t^4\,\Cov\bigl(\Var_\psi(\hat{H}_1), \Var_\psi(\hat{H}_2) \bigr) + O(t^6),
\end{eqnarray}
and therefore, we can rewrite the Pearson correlation coefficient in terms of the Hamiltonians as
\begin{eqnarray}
P(\hat{U}_1, \hat{U}_2) = P\bigr(\Var_\psi(\hat{H}_1), \Var_\psi(\hat{H}_2)\bigr) + O(t^2).
\end{eqnarray}
At the leading order, achieving $P = -1$ would force an almost-sure affine relation between the random functions $\Var_\psi(\hat{H}_1)$ and $\Var_\psi(\hat{H}_2)$: for all pure $\ket{\psi}$ and $a<0$,
\begin{eqnarray}
\Var_\psi(\hat{H}_2)= a\Var_\psi(\hat{H}_1) + b,
\label{eq:bq-affine-var}
\end{eqnarray}
because the correlation $-1$ with nonzero variances occurs if and only if one variable is an affine and strictly decreasing function of the other. Such a (state-independent) strictly decreasing affine dependence between the variance functions over the entire projective space is a highly rigid requirement. In particular, any Hamiltonian-level affine relation $\hat{H}_2=\alpha\hat{H}_1 + \beta\hat{\mathds{1}}$ would yield $\mathrm{Var}_\psi(\hat{H}_2)=\alpha^2 \mathrm{Var}_\psi(\hat{H}_1)$ (nonnegative slope), whereas Eq.~(\ref{eq:bq-affine-var}) would require $a<0$; hence the short-time condition for $P=-1$ is incompatible with nontrivial Hermitian generators~\cite{Ozawa2006,Geher2024}. We emphasize that this short-time discussion is included as an illustrative consistency check; the rigorous exclusion of $P=-1$ is established independently in {\bf Theorem~\ref{thm:no-minus-one}}. Hence, even in the infinitesimal-time limit a nonzero shared (channel-symmetric) component remains and prevents $P$ from reaching $-1$.

The readout design view implications from Sec.~\ref{sec:3} extend unchanged in spirit. Treating the survivals as features $(X_1, X_2)$ and forming the optimal linear contrast $C_\kappa = X_1 - \kappa X_2$ yields the same variance floor $\min_\kappa \Var(C_\kappa)=\Delta_1^2(1 - P^2)$ at $\kappa^\star=\Cov(X_1,X_2)/\Delta_2^2$, so the unattainability of $P=-1$ translates into an irreducible noise level in the difference channel. Across many-body Loschmidt echoes, the fundamental ceiling on ``anti-contrast'' is therefore imposed by \emph{unitary geometry}, not merely by pulse-area choices, and the sensible goal is to minimize the shared component between the two readouts while recognizing this unitary floor. Indeed, reaching $P=-1$ would require $\mathrm{Cov}(X_{\hat{U}_1}, X_{\hat{U}_2})^2=\mathrm{Var}(X_{\hat{U}_1}) \mathrm{Var}(X_{\hat{U}_2})$ (saturation of Cauchy--Schwarz), which by {\bf Theorem~\ref{thm:cs-pcc}} implies an affine dependence between the two survival-probability maps, ruled out for nontrivial unitaries by {\bf Theorem~\ref{thm:no-minus-one}}.

\section{Discussion}\label{sec:5}

Identifying survival probability $X_{\hat U}(\ket{\psi})=\abs{\bra{\psi}\hat{U}\ket{\psi}}^2$ as a canonical random variable on projective state space, we have shown that its Pearson correlation coefficient $P$ imposes a \emph{unitary-geometric} limit on how opposite two evolutions can be across an input ensemble. Within a Hilbert-space framework, $P$ inherits the Cauchy-Schwarz bound, with equality only when one random variable is an affine transform of the other; specializing this to survival probabilities and invoking the \emph{no quantum inversion} theorem ({\bf Theorem~\ref{thm:no-universal-inversion}}), we proved that the case $P=-1$ is \emph{forbidden} for any nontrivial unitary pair. Equivalently, no two unitaries have the survival probability maps that are point-wise complements on projective space. This closes a conceptual gap: Cauchy-Schwarz alone certifies $P \ge -1$ but is silent about the attainability of the lower edge; our result upgrades that edge to an \emph{unattainable, sharp bound} for unitary dynamics ({\bf Theorem~\ref{thm:no-minus-one}}). This is the first general and model-agnostic exclusion of the perfect anti-correlation of survival probabilities in finite dimension. We then instantiated the framework: in a single-qubit Bloch-sphere Ramsey model, we derived the closed form $P$ with a strict minimum $P_{\min} = -\frac{1}{2}$. In higher dimension, we showed that Haar/design moment identities express $P$ solely in terms of a handful of unitary invariants, guaranteeing $P > -1$ for any nontrivial pair. A short-time expansion linked this obstruction to a two-channel difference-measurement design scenario by tying the leading fluctuations of $X_{\hat{U}}$ to state-dependent energy variances, and a variance-rigidity argument showed that the strictly decreasing affine relation those variances would require for $P = -1$ cannot occur for nontrivial Hamiltonians. Taken together, these results establish a unitary-geometric floor on attainable anti-correlation.

This unitary-geometric floor has consequences for how we design and evaluate experiments. In feature space, the optimally weighted contrast $C_\kappa=X_1-\kappa X_2$ has the variance floor $\min_\kappa \Var(C_\kappa)=\Delta_1^2(1-P^2)$, so some residual shared leakage remains whenever $P > -1$---which, by our results, is always the case for nontrivial unitary pairs. This directly indicates that no ``purely difference-based'' parameter combination can be made noise-free by unitary control alone. These are practical constraints. They tell us what we \emph{can} tune (e.g., axis separation, interrogation time, and the input ensemble to decorrelate the relevant invariants) and what we \emph{cannot} achieve (e.g., a universal, point-wise bright-vs-dark tiling of state space). Because $P$ is insensitive to uniform contrast rescalings and concentrate quickly with sample size, they can serve as calibration-friendly figures of merit, for example, for benchmarking gates, tuning echo protocols, and certifying sensor working points.

We should note that the present work is formulated primarily for finite-dimensional (discrete-variable) Hilbert spaces, where the survival probabilities are averaged over the full projective state space admits a canonical unitarily invariant choice and enables the Haar-design moment identities used in Sec.~\ref{sec:4} and~{\bf \ref{appendix:A}}. In continuous-variable (CV) systems, on the other hand, the Hilbert space is infinite-dimensional and no normalizable unitarily invariant distribution over all pure states exists, so applying our PCC framework requires choosing a physically motivated (typically energy-constrained) input ensemble, such as coherent or squeezed/Gaussian states. The quantitative attainable minimum correlation can then become ensemble-dependent. Nevertheless, the key geometric obstruction underlying our main impossibility statement---that no nontrivial pair of unitaries can map every input to mutually orthogonal outputs---is Hilbert-space general; see~{\bf \ref{appendix:B}}. Many CV experiments also effectively operate within a finite-energy (truncated) subspace, in which case our finite-dimensional bounds apply directly on the operational subspace. A systematic analysis of ensemble-dependent bounds for Gaussian/CV sensing would be an interesting direction for a future study.

Looking ahead, our analysis points to experimental and theoretical directions. Experimentally, measuring $P$ over approximate projective designs provides a hardware-agnostic way to gauge how close a platform is to the unitary-geometric floor and to set quantitative targets for control optimization. Coupling the framework studied here with Fisher-space tomography would enable tracking of the elements of Fisher information matrix, turning our bounds into live diagnostics for sensitivity budgeting~\cite{Yu2021,Schreiber2025,Zhang2025}. Theoretically, a natural extension is to nonunitary dynamics (CPTP maps), replacing the survival probability, for example, by Uhlmann fidelity~\cite{Uhlmann1975} or channel survival~\cite{Knill2008}. Whether analogous quantum no-inversion obstructions persist, and how they interplay with noise and decoherence, are fertile questions with immediate metrological relevance. 

\section*{Acknowledgement}
JB thanks to Prof. Changsuk Noh for helpful comments. This work was supported by the Ministry of Science, ICT and Future Planning (MSIP) by the National Research Foundation of Korea (RS-2024-00432214, RS-2025-03532992, and RS-2023-NR119931) and the Institute of Information and Communications Technology Planning and Evaluation grant funded by the Korean government (RS-2019-II190003, ``Research and Development of Core Technologies for Programming, Running, Implementing and Validating of Fault-Tolerant Quantum Computing System''). This work is also supported by the Grant No.~K25L5M2C2 at the Korea Institute of Science and Technology Information (KISTI).

\appendix

\section{Covariance---Haar measure approach}\label{appendix:A}

In this appendix, we make use of the Haar measure on $U(d)$ to obtain explicit expressions for the expectation values and covariances between unitaries~\cite{CollinsSniady2006}. The basic strategy is to rewrite the state averages over $\ket{\psi}$ into group averages over $\hat{V} \in U(d)$ acting on a fixed reference state $\ket{0}$. This allows us to express all quantities in terms of moment operators associated with the Haar measure.

For a unitary $\hat{U} \in U(d)$, the mean $\E_{\psi}[X_{\hat{U}}]$ can be expressed as 
\begin{eqnarray}
\bar{f} := \E_{\psi}[X_{\hat{U}}] &=& \int d\psi \abs{\bra{\psi}\hat{U}\ket{\psi}}^2 = \int d \psi \bra{\psi}^{\otimes 2} \bigl( \hat{U}^{\dagger} \otimes \hat{U} \bigr) \ket{\psi}^{\otimes 2} \nonumber \\
&=& \bra{00}  \left( \int \hat{V}^{\dagger \otimes 2} (\hat{U}^{\dagger} \otimes \hat{U}) \hat{V}^{\otimes 2} d\mu(\hat{V}) \right) \ket{00},
\end{eqnarray}
which represents the average survival probability associated with $\hat{U}$. Here, we employ the standard tensor-trick, rewriting an arbitrary pure state as $\ket{\psi} = \hat{V}\ket{0}$ with $\hat{V} \in U(d)$ and averaging over the Haar measure on $U(d)$.  

Applying the same trick, $\E_{\psi}[X_{\hat{U}_{1}} X_{\hat{U}_{2}}]$ can be expressed as
\begin{eqnarray}
\bar{d}_{1, 2} &:=& \E_{\psi}[X_{\hat{U}_{1}} \, X_{\hat{U}_{2}}] = \int d\psi \abs{\bra{\psi}\hat{U}_{1}\ket{\psi}}^2 \abs{\bra{\psi}\hat{U}_{2}\ket{\psi}}^2, \nonumber \\
 &=& \int d\psi \bra{\psi}^{\otimes 4} \hat{U}_{1}^{\dagger} \otimes \hat{U}_{1} \otimes \hat{U}_{2}^{\dagger} \otimes \hat{U}_{2} \ket{\psi}^{\otimes 4}  \nonumber \\
 &=& \bra{0000} \left( \int \hat{V}^{\dagger \otimes 4} \bigl( \hat{U}_{1}^{\dagger} \otimes \hat{U}_{1} \otimes \hat{U}_{2}^{\dagger} \otimes \hat{U}_{2} \bigr)   \hat{V}^{\otimes 4}  d\mu(\hat{V}) \right) \ket{0000}.
\end{eqnarray}
It turns out that the integrals $\bar{f}$ and $\bar d_{1,2}$ are nothing but the second and fourth moment operators of the Haar measure on $U(d)$, respectively. The evaluation of these quantities thus reduces to the unitary $t$-designs~\cite{Gross2007,Ambainis2007} and symmetric group representation theory~\cite{Goodman2009}. For a detailed discussion of Haar moments and their use in this context, see Refs.~\cite{Zhang2014,Ragone2022}.  

In order to compute the Haar integrals explicitly, we recall the following propositions which formulates the general expressions for the $k$-th moment operators and the corresponding trace formulas involving permutations~\cite{Cho2025}.
\begin{proposition}
The contraction of $k$-th moment of Haar measure, $\mathcal{M}_{\mu_H}^{(k)}(\hat{\mathcal{O}}) := \E_{\hat{U} \sim \mu_H} [\hat{U}^{\otimes k} \hat{\mathcal{O}} \hat{U}^{\dagger \otimes k}]$ can be obtained 
\begin{eqnarray}
\fl \bra{0}^{\otimes k} \E_{\mu_H \sim \hat{U}} [\hat{U}^{\otimes k}\hat{\mathcal{O}}\hat{U}^{\dagger \otimes k}] \ket{0}^{\otimes k}  = \frac{1}{d(d+1)\cdots (d+k-1)}  \left(\sum_{\pi \in S_k} \tr(\hat{V}_d^{\dagger}(\pi) \hat{\mathcal{O}}) \right),
\end{eqnarray}
where $\hat{V}_d(\pi)$ is called the permutation operator of symmetric group $S_k$, defined as
\begin{eqnarray}
\hat{V}_d(\pi) = \sum_{i_1, \cdots, i_k \in [d]^k} \ket{i_{\pi^{-1}(1)} , \cdots, i_{\pi^{-1}(k)}} \bra{i_1, \cdots, i_k}.
\end{eqnarray}
\end{proposition}
To evaluate these expressions in practice, one requires formulas for the traces of operators contracted with permutation operators. The following proposition provides such a formula.
\begin{proposition}
Let $\pi \in S_k$, and let $\hat{A}_1, \dots, \hat{A}_k$ be operators represented as $d \times d$ matrices over $\mathbb{C}$. Let $\hat{V}_d(\pi)$ denote the permutation operator acting on the tensor product space $(\mathbb{C}^d)^{\otimes k}$. Then,
\begin{eqnarray}
\tr \bigl( \hat{A}_1 \otimes \cdots \otimes \hat{A}_k \hat{V}_d(\pi) \bigr) = \prod_{c \in \pi} \tr \left( \prod_{m=0}^{k_c -1} \hat{A}_{c^{-m} (l_c)} \right) 
\end{eqnarray}
where $\pi = \{ c_1, \dots, c_r \}$ is the disjoint cycle decomposition of $\pi$, each cycle $c = (l_1, \dots, l_{k_c})$ has length $k_c = \abs{c}$, and $l_c$ is an arbitrary reference element of cycle $c$. The notation $c^{-m}(l_c)$ denotes the $m$-th inverse image of $l_c$ under the cycle $c$.
\end{proposition}

With these two propositions, we are in a position to evaluate the Haar integrals. In particular, applying the $k=2, 4$ cases gives the following closed-form expressions for $\bar{f}_{j}$ and $\bar{d}_{1, 2}$:
\begin{eqnarray} 
&& \bar{f}_{j} = \frac{1}{d(d+1)} \left( d + \abs{\tr(\hat{U}_{j})}^2 \right) \quad (j=1,2), \nonumber \\
&& \bar{d}_{1, 2} = \frac{1}{d(d+1)(d+2)(d+3)} \Big[ d(d+4) + (d+4) \left( \abs{\tr(\hat{U}_{1})}^2 + \abs{\tr(\hat{U}_{2})}^2 \right)  \nonumber \\
&& \qquad
+ \left( \abs{\tr(\hat{U}_{1})}^2 \abs{\tr(\hat{U}_{2})}^2+ \abs{\tr(\hat{U}_{1} \hat{U}_{2})}^2 + \abs{\tr(\hat{U}_{1} \hat{U}_{2}^{\dagger})}^2 \right)  \nonumber \\
&& \qquad
+ 2 \text{Re} \bigl[ \tr(\hat{U}_{1} \hat{U}_{2}) \tr(\hat{U}_{1}^{\dagger} ) \tr(\hat{U}_{2}^{\dagger}) \bigr]+ 2 \text{Re} \bigl[ \tr(\hat{U}_{1} \hat{U}^{\dagger}_{2} ) \tr(\hat{U}_{1}^{\dagger}) \tr(\hat{U}_{2}) \bigr] \nonumber \\
&& \qquad
 +2 \text{Re} \bigl[ \tr(\hat{U}_{1} \hat{U}_{2} \hat{U}_{1}^{\dagger} \hat{U}_{2}^{\dagger})\bigr) \bigr] \Big].
\label{bard}
\end{eqnarray}
From these formulas, we immediately obtain closed-form expressions for the covariance and variance. Recall that the covariance is defined as
\begin{eqnarray}
\mathrm{Cov}(X_{\hat{U}_{1}},X_{\hat{U}_{2}}) = \bar{d}_{1, 2} - \bar{f}_{1} \bar{f}_{2},
\label{cov}
\end{eqnarray}
while the variance is obtained from 
\begin{eqnarray}
\Delta_j^2 = \bar{d}_{j} - \bar{f}_j^2, \qquad \bar{d}_{j} := \bar{d}_{j,j}. \label{std}
\end{eqnarray}
A direct evaluation gives
\begin{eqnarray}
\fl \Delta_j^2 = \frac{2d(d+3) + 4(d+2)\abs{\tr(\hat{U}_{j})}^2 + \abs{\tr(\hat{U}_{j}^2)}^2 + \abs{\tr(\hat{U}_{j})}^4 + 2\text{Re}\bigl( \tr(\hat{U}_{j}^2)\, \tr(\hat{U}_{j}^{\dagger})^2 \bigr)}{d(d+1)(d+2)(d+3)} - \bar{f}_{j}^2. \nonumber \\
\fl \Cov(X_{\hat{U}_{1}}, X_{\hat{U}_{2}}) = \frac{1}{d(d+1)(d+2)(d+3)}  \Big[ -4d + 4d(d+1)(\bar{f}_{1} + \bar{f}_{2}) \nonumber \\
\quad
  - 2(2d+3)d(d+1)\bar{f}_{1} \bar{f}_{2} + \abs{\tr(\hat{U}_{1} \hat{U}_{2})}^2 + \abs{\tr(\hat{U}_{1} \hat{U}_{2}^\dagger)}^2 \nonumber \\
\quad
  + 2 \text{Re} \bigl[ \tr(\hat{U}_{1} \hat{U}_{2}) \tr(\hat{U}_{1}^{\dagger} ) \tr(\hat{U}_{2}^{\dagger}) \bigr]+ 2 \text{Re} \bigl[ \tr(\hat{U}_{1} \hat{U}^{\dagger}_{2} ) \tr(\hat{U}_{1}^{\dagger}) \tr(\hat{U}_{2}) \bigr] \nonumber \\
\qquad
   +2 \text{Re} \bigl[ \tr(\hat{U}_{1} \hat{U}_{2} \hat{U}_{1}^{\dagger} \hat{U}_{2}^{\dagger})\bigr) \bigr] \Big].
\end{eqnarray}
With the explicit form of the unitaries $\hat{U}_{1}, \hat{U}_{2}$, one can derive concrete expressions for the covariance and Pearson correlation coefficient. In particular, for $SU(2)$, a qubit subject to two coherent controls, $\hat{U}_j=\exp\left(-\frac{i}{2} \theta_j \mathbf{n}_j \cdot \boldsymbol{\sigma} \right)$,  the parametrization in terms of Pauli matrices and their exponentials leads to especially simple closed forms.

\subsection{$SU(2)$: Bloch sphere via Haar measure} 

We now illustrate these formulas concretely for the simplest nontrivial case, namely $SU(2)$. Any unitary $\hat{U}_{j} \in SU(2)$ can be parametrized as
\begin{eqnarray}
\hat{U}_{j} = \exp\left( -\frac{i}{2} \theta_j \mathbf{n}_j \cdot \boldsymbol{\sigma} \right) = \cos \left(\frac{\theta_j}{2} \right) \hat{\mathds{1}} - i \sin\left(\frac{\theta_j}{2}\right)\mathbf{n}_j \cdot \boldsymbol{\sigma},
\end{eqnarray}
where $\mathbf{n}_{j} \in \mathbb{R}^3$ is a unit Bloch axes and $\boldsymbol{\sigma}=(\hat{\sigma}_x, \hat{\sigma}_y, \hat{\sigma}_z)^T$ are the Pauli matrices forming the basis of $SU(2)$.  

To evaluate the covariance and Pearson correlation coefficient, one needs explicit trace identities involving these unitaries. Direct computation gives
\begin{eqnarray}
&& \tr(\hat{U}_j) = 2 \cos\left(\frac{\theta_j}{2}\right), \\
&& \tr(\hat{U}_{1} \hat{U}_{2}) = 2 \cos\left(\frac{\theta_{1}}{2}\right) \cos\left(\frac{\theta_{2}}{2}\right) - 2 (\mathbf{n}_{1} \cdot \mathbf{n}_{2}) \sin\left(\frac{\theta_{1}}{2}\right) \sin\left(\frac{\theta_{2}}{2}\right), \\
&& \tr(\hat{U}_{1} \hat{U}_{2}^{\dagger}) = 2 \cos\left(\frac{\theta_{1}}{2}\right) \cos\left(\frac{\theta_{2}}{2}\right) + 2 (\mathbf{n}_{1} \cdot \mathbf{n}_{2}) \sin\left(\frac{\theta_{1}}{2}\right) \sin\left(\frac{\theta_{2}}{2}\right), \\ 
&& \tr(\hat{U}_{1} \hat{U}_{2} \hat{U}_{1}^{\dagger} \hat{U}_{2}^{\dagger}) = 2 - 2 \bigl(1- ( \mathbf{n}_{1} \cdot \mathbf{n}_{2} )^2 \bigr) \bigl(1-\cos\theta_{2}\bigr) \sin^2\left(\frac{\theta_{1}}{2}\right).
\end{eqnarray}
Using these identities, we obtain
\begin{eqnarray}
&& \bar{f}_{j} = \frac{1}{3}\left( 2 + \cos\theta_{j} \right), \nonumber \\
&& \Delta_{j}^2 = \Cov(X_{\hat{U}_{j}}, X_{\hat{U}_{j}}) = \frac{4}{45} \sin^4 \frac{\theta_{j}}{2} = \frac{1}{5} \bigl( 1 - \bar{f}_{j} \bigr)^2, \nonumber \\
&& \Cov(X_{\hat{U}_{1}}, X_{\hat{U}_{2}}) = \frac{1}{10} \bigl(-1 + 3 ( \mathbf{n}_{1} \cdot \mathbf{n}_{2} )^2 \bigr) \bigl(1-\bar{f}_{1} \bigr) \bigl(1 - \bar{f}_{2}\bigr).
\end{eqnarray}
Therefore, the Pearson correlation coefficient for $SU(2)$ admits the closed form
\begin{eqnarray}
P(\hat{U}_{1}, \hat{U}_{2}) = \frac{\Cov(X_{\hat{U}_{1}},  X_{\hat{U}_{2}})}{\sqrt{\Delta_{1}^2} \sqrt{\Delta_{2}^2}} = \frac{3 \cos^2 \delta -1}{2}.
\end{eqnarray}
where $\mathbf{n}_{1} \cdot \mathbf{n}_{2} = \cos \delta$; hence, the correlation is bounded as $P(\hat{U}_{1}, \hat{U}_{2}) \in \bigl[-\frac{1}{2}, 1 \bigr]$ with $\cos^2 \delta \in [0,1]$.

\section{Continuous-variable protocols and infinite-dimensional Hilbert spaces}\label{appendix:B}

This appendix clarifies (i) which parts of our analysis rely on finite dimension and the existence of a canonical unitarily invariant state ensemble, and (ii) how an analogous PCC-based comparison can be formulated for continuous-variable (CV) protocols once a physically motivated input ensemble is specified.

\subsection{Why the finite-dimensional ``uniform over all pure states'' average is special}

In finite dimension, the projective space $\mathbb{CP}^{d-1}$ is compact and carries a unique unitarily invariant probability measure, which provides a canonical notion of ``scanning all pure inputs uniformly.'' In infinite-dimensional CV Hilbert spaces, the unitary group is not compact and there is no normalizable unitarily invariant probability measure over all pure states that plays an analogous role. Therefore, a CV analogue of our device-agnostic state averaging must be tied to a certain physical constraint, most commonly an energy (mean-photon-number) constraint, and hence to a specified ensemble, such as coherent states, squeezed/Gaussian states, or other experimentally relevant families.

\subsection{Relation to our main impossibility result and to effective truncations}

In fact, our {\bf Theorem~\ref{thm:no-minus-one}} implies that the point-wise perfect opposition condition (mutual orthogonality of outputs for every input state) is excluded in any Hilbert space, including CV settings. This supports the qualitative message that a universal bright-versus-dark tiling of the entire state space cannot be achieved by nontrivial unitary dynamics.

What is finite-dimensional specific in our main text is the stronger statement formulated via the canonical full projective-state ensemble: the proof of {\bf Theorem~\ref{thm:no-minus-one}} uses (i) the equality condition for PCC ($P=-1$ implies an affine relation almost surely with respect to the chosen ensemble) and (ii) the fact that this would enforce a point-wise complementarity structure incompatible with {\bf Theorem~\ref{thm:no-universal-inversion}} and {\bf Corollary~\ref{cor:no-perfect-complement}}. In a CV protocol where one averages only over a restricted ensemble, a PCC value close to $-1$ would correspond to an almost-sure affine relation on that restricted support, and a separate analysis is required to decide whether $P = -1$ is attainable.

At the same time, many CV experiments operate effectively within a finite-energy subspace or an encoded finite-dimensional manifold. In such cases, our finite-dimensional results apply directly to the operational subspace. A simple continuity bound illustrates this point. Let $\hat{\Pi}$ be the projector onto an operational subspace $\mathcal{K}$ and let $p:=\bra{\psi}\hat{\Pi}\ket{\psi} \ge 1-\varepsilon$. Then, let us define a normalized truncated state $\ket{\psi_\star}:=\hat{\Pi}\ket{\psi}/\sqrt{p}$. For any unitary $\hat{U}$,
\begin{eqnarray}
\abs{\bra{\psi}\hat{U}\ket{\psi}}^2 - \abs{\bra{\psi_\star}\hat{U}\ket{\psi_\star}}^2 \le 4\sqrt{1 - \abs{\braket{\psi}{\psi_\star}}^2} =4\sqrt{\varepsilon},
\label{eq:B6}
\end{eqnarray}
where we used the trace-distance bound $\abs{\tr{[\hat{U}(\hat{\rho}-\hat{\rho}')]}} \le 2D(\hat{\rho}, \hat{\rho}')$ (valid for $\|\hat{U}\|=1$) together with the distance $D(\ketbra{\psi}{\psi}, \ketbra{\psi_\star}{\psi_\star})
=\sqrt{1 - \abs{\braket{\psi}{\psi_\star}}^2}$.
Thus, when the inputs have small leakage outside $\mathcal{K}$, the survival probabilities---and therefore, ensemble means/covariances computed from them---are stable under truncation, and the finite-dimensional analysis provides a controlled approximation on the operational subspace.

\subsection{B.5. Example: coherent-state ensemble under phase shifts}

To illustrate ensemble dependence concretely, consider a single bosonic mode with the number operator $\hat{n}=\hat{a}^\dagger \hat{a}$ and phase shift $\hat{U}(\theta)=e^{-i\theta \hat{n}}$. For a coherent state $\ket{\alpha}$,
\begin{eqnarray}
\hat{U}(\theta)\ket{\alpha} &=& \ket{\alpha e^{-i\theta}}, \nonumber \\
X_\theta(\alpha) &=& \abs{\bra{\alpha}\hat{U}(\theta)\ket{\alpha}}^2 = \abs{\braket{\alpha}{\alpha e^{-i\theta}}}^2 = e^{-4|\alpha|^2\sin^2(\theta/2)}.
\label{eq:B7}
\end{eqnarray}
Now, choose a simple energy-constrained ensemble: an isotropic complex Gaussian distribution
\begin{eqnarray}
d\mu_{\bar{n}}(\alpha) = \frac{d^2\alpha}{\pi \bar{n}}e^{-\frac{|\alpha|^2}{\bar{n}}},
\label{eq:B8}
\end{eqnarray}
for which $\mathbb{E}_{\mu_{\bar{n}}}[|\alpha|^2]=\bar{n}$. Then, the gaussian integration yields the closed-form moments
\begin{eqnarray}
\mathbb{E}_{\mu_{\bar{n}}}[X_\theta] &=& \frac{1}{1+4\bar{n}\sin^2(\theta/2)}, \nonumber \\
\mathbb{E}_{\mu_{\bar{n}}}[X_\theta^2] &=& \frac{1}{1+8\bar{n}\sin^2(\theta/2)}, \nonumber \\
\mathbb{E}_{\mu_{\bar{n}}}[X_{\theta_1}X_{\theta_2}] &=& \frac{1}{1+4\bar{n}\left(\sin^2(\theta_1/2)+\sin^2(\theta_2/2)\right)}.
\label{eq:B9}
\end{eqnarray}
From the above Eq.~(\ref{eq:B9}), one obtains $\mathrm{Var}_{\mu_{\bar{n}}}(X_\theta)$ and $\mathrm{Cov}_{\mu_{\bar{n}}}(X_{\theta_1},X_{\theta_2})$, and hence the ensemble-dependent correlation $P_{\mu_{\bar{n}}}(\hat{U}(\theta_1),\hat{U}(\theta_2))$. Notably, in this example the covariance is nonnegative because both $X_{\theta_1}(\alpha)$ and $X_{\theta_2}(\alpha)$ are monotone decreasing functions of $|\alpha|^2$ under the same ensemble. Therefore, one cannot achieve the negative correlation, let alone $-1$, by comparing mere phase shifts on this coherent-state ensemble. This illustrates that CV “anti-contrast” performance depends crucially on both the chosen family of unitaries and the physically relevant input ensemble.

\section*{References}

\bibliographystyle{iop}

\end{document}